\documentclass[pra,twocolumn,preprintnumbers,showpacs]{revtex4}
\usepackage{graphicx}
\begin{document}

\title{Simple and efficient generation of gap solitons in
Bose-Einstein condensates}

\author{Micha\l{} Matuszewski$^{1,2}$, Wieslaw Kr\'olikowski$^{2}$,
Marek Trippenbach$^{1,3}$, and Yuri S. Kivshar$^{4}$}

\affiliation{$^1$Institute for Theoretical Physics, Warsaw
University, Ho\.{z}a 69, PL-00-681 Warsaw, Poland \\
$^2$Laser Physics Centre, Research School of Physical Sciences and
Engineering, Australian National
University, Canberra ACT 0200, Australia \\
$^3$Soltan Institute for Nuclear Studies, Ho\.{z}a 69, PL-00-681
Warsaw, Poland\\
$^4$Nonlinear Physics Centre and ARC Centre of Excellence for
Quantum-Atom Optics, Research School of Physical Sciences and
Engineering, Australian National University, Canberra ACT 0200,
Australia}

\begin{abstract}
We suggest an efficient method for generating matter-wave gap
solitons in a repulsive Bose-Einstein condensate, when the gap
soliton is formed from a condensate cloud in a harmonic trap after
turning on a one-dimensional optical lattice. We demonstrate
numerically that this approach does not require preparing the
initial atomic wavepacket in a specific state corresponding to the
edge of the Brillouin zone of the spectrum, and losses that occur
during the soliton generation process can be suppressed by an
appropriate adiabatic switching of the optical lattice.
\end{abstract}

\pacs{03.75.Lm, 05.45.Yv}

\maketitle

\section{Introduction}

Recent experimental developments in the fields of nonlinear optics
and Bose-Einstein condensation provide unique opportunities for the
studies of many nonlinear phenomena. Moreover, due to a formal
similarity of the Gross-Pitaevskii equation for the dynamics of
matter waves and the nonlinear Schr\"odinger equation for light
propagating in a nonlinear Kerr medium, it is possible to
interchange and transfer many interesting ideas between these two
fields. One of them is the generation of nonspreading localized
wavepackets usually called {\em optical solitons} or {\em
matter-wave solitons}~\cite{KivsharAgrawal}.

Bright solitons, which are quite common objects in different
problems of nonlinear optics, have been only recently demonstrated
in two independent experiments with matter
waves~\cite{Khaykovich,Strecker}. In this case, the wavepacket
spreading due to dispersion is compensated by an attractive
interaction between atoms. Both optical and matter-wave solitons can
also be formed due to an interplay between nonlinearity and
periodicity~\cite{Christodoulides,Ostrovskaya,Ostrovskaya2}.

In optics, the presence of a periodically varying refractive index
of a medium results in a bandgap structure of the transmission
spectrum of light. This subsequently affects the propagation of
optical beams (pulses), which is determined by the diffraction
(dispersion) curves of the transmission bands and forbidden gaps of
the wave spectrum. It was demonstrated that periodic structures can
modify dramatically the wave diffraction~\cite{Eisenberg}, and
therefore they introduce a novel way for controlling self-action of
light in nonlinear periodic media~\cite{Fleischer,Sukhorukov}. In
particular, if the refractive index decreases with the light
intensity due to a nonlinearity-induced response of the material,
the beam normally experiences broadening due to self-defocusing.
However, in periodic photonic structures, the same type of
defocusing nonlinearity can lead to the beam
localization~\cite{Kivshar_OL}. For deep periodic modulations, this
effect should result in the formation of discrete solitons. Indeed,
as was shown both numerically and experimentally in
Ref.~\cite{Krol}, increasing the depth of the refractive index
modulation leads to a sharp crossover between the continuous and
discrete properties of the lattice. It is manifested in a specific
type of  light localization and gap-soliton generation which become
possible for a self-defocusing medium only in case of a discrete
model.

Analogous effect of the gap-soliton generation in the regime of
repulsive interaction was observed for the Bose-Einstein
condensates. This first experimental observation~\cite{Oberthaler}
was reported for $^{87}$Rb atoms in a weak optical lattice. In this
case, the anomalous dispersion of the atomic cloud is realized at
the edge of the Brillouin zone; if the coherent atomic wave packet
is prepared in the corresponding state, a matter-wave gap soliton is
formed inside the spectral gap. However, it took a considerable
experimental effort to generate the initial state of the
Bose-Einstein condensate in a proper way, by introducing an
acceleration to the optical lattice and other
tricks~\cite{Burger,Oberthaler}.

In this paper, we reveal that the matter-wave gap solitons can be
generated by a simpler and straightforward method that allows to
create stable, long-lived gap solitons of a repulsive Bose-Einstein
condensate in a one-dimensional optical lattice. In our scenario,
the gap soliton is formed from an initial ground state of a harmonic
trap after turning on the optical lattice, due to the cloud
relaxation rather than as a result of preparing the initial atomic
wavepacket at a certain point of the band gap structure. In
addition, we demonstrate that losses during this process
can be suppressed by an appropriate adiabatic switching of the
optical lattice.

The paper is organized as follows. In Sec.~II we introduce our
theoretical model. Then, in Sec.~III we present two methods for
generating the matter-wave gap solitons and demonstrate their
efficiency in numerical simulations. The coupled-mode theory for
describing the self-trapping mechanism is introduced in Sec.~IV
where we discuss the essential physics of the localization process.
Finally, Sec.~V concludes the paper.

\section{Theoretical model}

We analyze our matter-wave system in the framework of the
three-dimensional mean-field Gross-Pitaevskii equation for the
macroscopic wave function $\Psi({\bf r}, t)$, written in the
standard form,
\begin{equation}
i \hbar\frac{\partial \Psi}{\partial t}=\left(
-\frac{\hbar^2}{2m}\Delta +U(\mathbf{r},t)+ \frac{4 \pi a
\hbar^2}{m}|\Psi|^{2}\right) \Psi, \label{GPE}
\end{equation}
where for the repulsive condensate the scattering length is positive
($a>0$).  We assume that the external potential $U(\mathbf{r},t)$
consists of two parts,  a harmonic trap and one-dimensional optical
lattice, i.e.
\[
{U}(\mathbf{r},t)={\varepsilon}(t) \sin^2 \left(\frac{\pi
z}{d}\right) + \frac{m}{2}\left[{\omega}_{\perp}^{2}\varrho ^{2} +
f(t){\omega}_{z}^{2}z ^{2}\right],
\]
where $d=\lambda/2$ is the period of the optical lattice potential,
$\rho = (x^2+y^2)^{1/2}$ is the two-dimensional radial coordinate,
and the cylindrically-symmetric harmonic trap is characterized by
the longitudinal and transverse frequencies, ${\omega}_{z}$ and
${\omega}_{\perp}$, respectively. We also introduce the depth
${\varepsilon}$ and period $\lambda$ of an optical lattice. Temporal
variations of the potential are described by two functions, $f_1(t)$
and $f_2(t)$, which  account for the adiabatic
processes of turning off the longitudinal part of the harmonic trap
and turning on the optical lattice over the stationary condensate.
More details about this tuning will follow in the next section.

In our numerical calculations, we use both
three-dimensional model and (more often) quasi-one-dimensional
reduced model, in which we assume that the
transverse harmonic confinement is so tight that no transverse
excitations occur~\cite{Gerlitz,OberthalerPRA}. As a result of this
assumption, the three-dimensional wave function resembles the
transverse ground state of a harmonic potential in the transverse
cross-section, and the nonlinear dynamics can be described by a
one-dimensional Gross-Pitaevskii equation after rescaling the
nonlinear coefficient, $a_{1D}= (m \omega_{\perp}/2 \pi \hbar)a $.

\section{Numerical Results and Discussions}

In our numerical simulations, we suggest and analyze systematically
{\em two methods} for generating matter-wave gap solitons in the
repulsive Bose-Einstein condensates loaded in a shallow
one-dimensional optical lattice.

\begin{figure}[tbp]
\includegraphics[width=8.5cm]{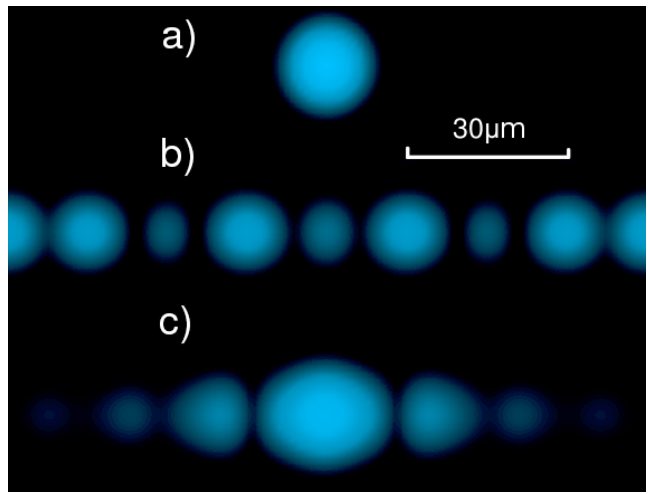}
\caption{(color online) Generation of a matter-wave gap soliton.
Shown are the density snapshots for (a,b) $N=200$ atoms at the input
($t=0$) and after expansion ($t=0.4$s), respectively, and (c) for
$N=500$ atoms, after expansion at $t=1.4$s. The lattice parameters
are: $\varepsilon=2.3 E_{\mathrm{recoil}}$ and period  $15 \mu$m. }
\label{columns}
\end{figure}

\subsection{The first method}

 In the first method,  we start from the $^{87}$Rb condensate (atomic mass $86.9$ amu and
the scattering length $5.3\times 10^{-9}$ m) as an eigenstate of a
cylindrically symmetric harmonic trap with the frequencies
$\omega_{z0}=2\pi\times 40$Hz and $\omega_{\perp}=2\pi\times 50$Hz.
Then, we instantaneously turn off the longitudinal trapping
potential leaving the perpendicular harmonic potential on and
simultaneously switch on, in the same direction, a one-dimensional
optical lattice with the spatial period $\lambda=30\mu$m. For this
lattice we define the recoil energy $E_{\mathrm{recoil}}=\hbar^2(2
\pi/\lambda)^2/(2m)$. Such a lattice can be created by interfering
two laser beams propagating at a small
angle~\cite{Sukhorukov,Neshev}. For the wavelength of $783$ nm, used
in Ref.~\cite{Oberthaler}, this angle would be equal to $2.3$
degrees. After an abrupt change of the potential, a
quasi-one-dimensional dynamics takes place, in which atoms slowly
tunnel to the neighboring lattice sites, while the confinement in
the direction perpendicular to the optical lattice remains
unaffected. As is shown in Figs.~\ref{columns}(a-c), depending on
the initial number of atoms, we observe asymptotically either
expansion of the condensate over more and more lattice sites [see
Fig.~\ref{columns}(b)] or condensate stabilization accompanied by
the formation of a spatially-localized state occupying just a few
lattice sites [see Fig.~\ref{columns}(c)]. We checked that there
exists a threshold in the initial number of atoms (with all the rest
of the system parameters fixed) above which a stable spatially
localized state is generated in the optical lattice.

\begin{figure}[tbp]
\includegraphics[width=8.5cm]{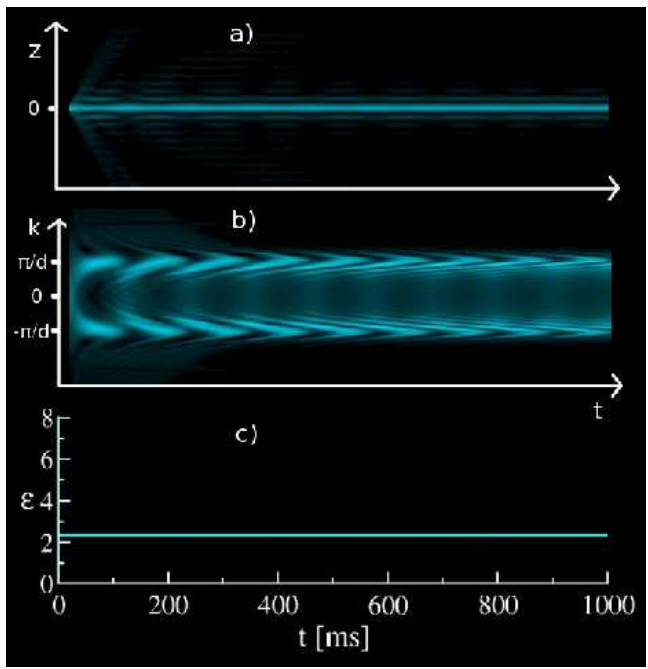}
\caption{(color online) (a) Temporal evolution (from left to right)
of the axial density of the condensate [see Fig.~\ref{columns}(c)].
(b) Evolution shown in the Fourier space. (c) Temporal variation of
the depth of the optical lattice (cf. Fig.~5
below).}\label{adibatic2}
\end{figure}

\begin{figure}[tbp]
\includegraphics[width=8.5cm]{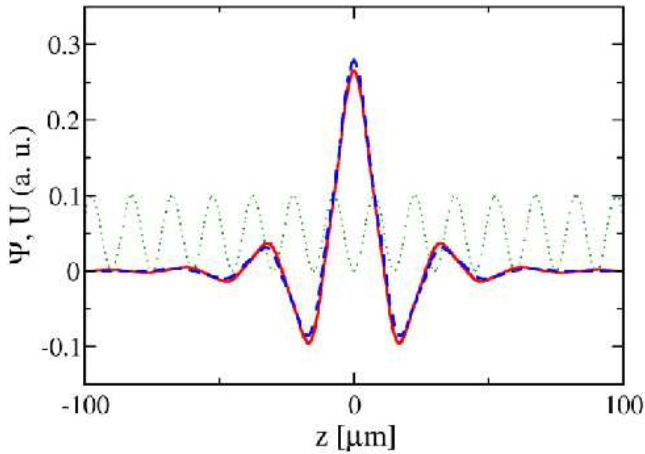}
\caption{(color online) Wavefunction of the matter-wave gap soliton
of Fig.~\ref{columns}(c), as predicted by the one-dimensional
Gross-Pitaevskii equation (dashed) and the full three-dimensional
model (solid); Dotted line shows the lattice potential.}\label{profil}
\end{figure}

These results are summarized in Figs.~\ref{adibatic2}(a-c), where we
show the profile of the soliton wavefunction and the corresponding
dynamics in the Fourier space. In the Fourier space, a clear
localization at the edges of the first band occurs. To understand
importance and specific features of this localization process, we
notice that the evolution of a condensate in a one-dimensional
optical lattice can be described within the effective mass
approximation
\begin{equation}
i \hbar\frac{\partial \psi}{\partial t}= -\frac{\hbar^2}{2m_{\rm
eff}}\frac{\partial^2 \psi}{\partial z^2} + \frac{4 \pi a_{\rm eff}
\hbar^2}{m}|\psi|^{2}\psi, \label{GPEef}
\end{equation}
where $\psi(z,t)$ is {\em an envelope} of the condensate
wavefunction at the edge of the Brillouin zone of the spectrum. The
localized state can be supported only if the sign of the effective
mass $m_{\rm eff}$ is opposite to the sign of the scattering length
$a_{\rm eff}$. Since for a repulsive condensate $a_{\rm eff}>0$, the
effective mass in this case has to be negative (anomalous
diffraction). This condition is indeed fulfilled for the matter
waves close to the edge of the Brillouin zone. In Sec.~V we present
a simple model that allows to describe this kind of the localization
process.

\begin{figure}[tbp]
\includegraphics[width=8.5cm]{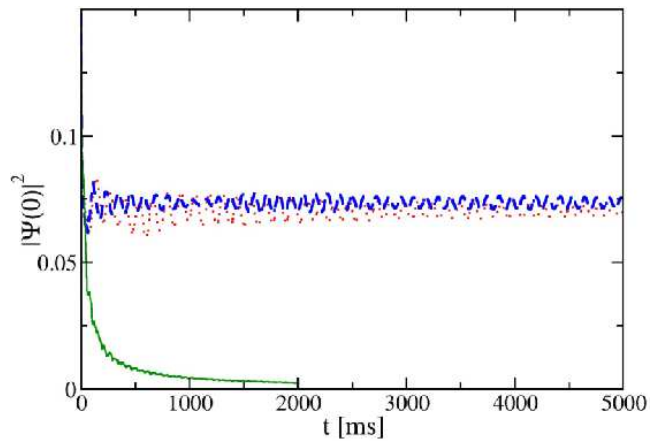}
\caption{(color online) Evolution of the peak amplitude of the
atomic cloud in the lattice for (a) $N=200$ atoms (solid), (b,c)
$N=500$ atoms in the one-dimensional (dashed) and three-dimensional
(dotted, rescaled) models, respectively.}\label{evol}
\end{figure}

In Fig.~\ref{profil}, we show the wavefunction profile for the case
of $N=500$ atoms at $t=1.4$s, when the gap soliton is generated, as
shown in Fig.~\ref{columns}(c). The phase structure of the resulting
state corresponds to the edge of the first band of the Brillouin
zone~\cite{Louis}. In addition, we compare the results obtained from
the integration of the one- and full three-dimensional
Gross-Pitaevskii models, and notice that the condensate dynamics
looks practically the same in both the cases. With a good accuracy,
we can therefore assume that the condensate remains in the ground
state of the transverse harmonic potential that is left on all the
time. Also, in Fig.~\ref{evol} we show the evolution of the peak
amplitude of the atomic cloud that fully supports our observation.
As mentioned above, for $N=200$ atoms we predict overall spreading
of the wave packet and the peak amplitude vanishes gradually. For
larger number of the input atoms (e.g., $N=500$), the numerical
simulations of both one- and three-dimensional models show the
stabilization effect and both approaches predict almost the same
final distribution of atoms on a few neighboring lattice sites. In
earlier studies~\cite{Oberthaler,OberthalerPRA}, the generated gap
soliton was unstable due to the excitation of transverse modes of
the harmonic trap. Here, we suggest applying the optical lattice
with the period comparable to the transverse extent of the
condensate. Such a configuration (i.e. a sparser lattice) has much
more favorable ratio of the energies of the transverse and
longitudinal modes, and the suppression of the instability is
observed.

\begin{figure}[tbp]
\includegraphics[width=8.5cm]{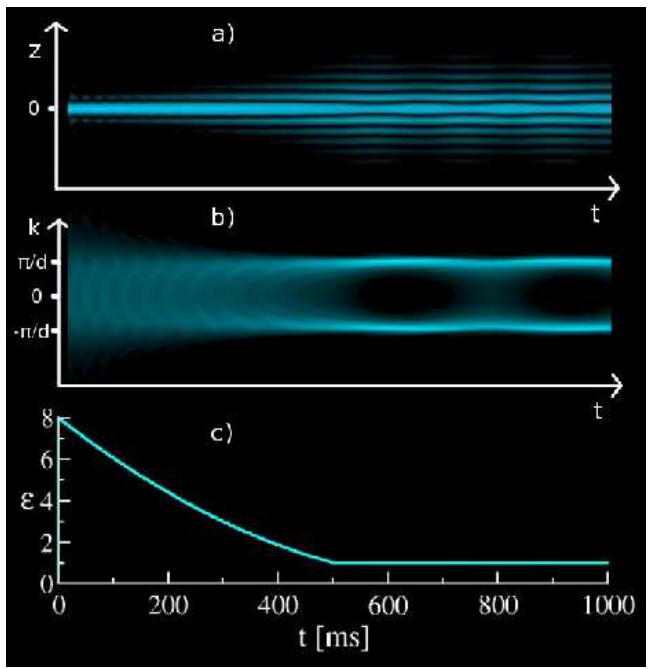}
\caption{(color online) Top: Temporal evolution (from left to right)
of the axial density of the condensate in the case of adiabatic
reduction of the lattice potential from $\varepsilon=8
E_{\mathrm{recoil}}$ to $\varepsilon=E_{\mathrm{recoil}}$ during
first $t=500$ms. Middle: Evolution shown in Fourier space.
Bottom: Temporal variation of the depth of the
optical lattice.}\label{adibatic}
\end{figure}

\subsection{The second method}

\begin{figure}[tbp]
\includegraphics[width=8.5cm]{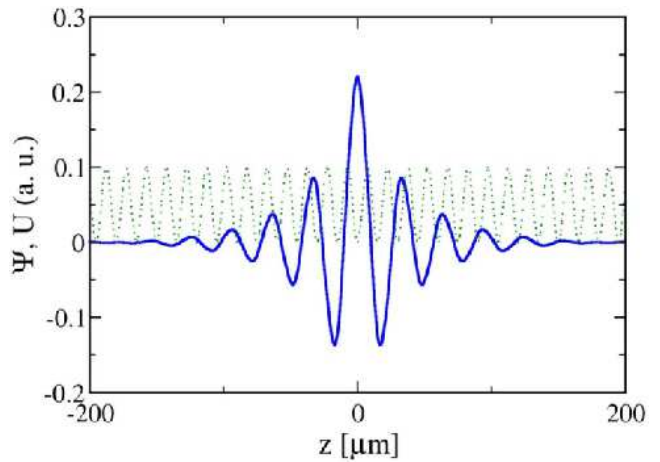}
\caption{(color online) Profile of the condensate wavefunction at
$t=1$s for the gap soliton generation shown in Fig.~\ref{adibatic}.
Dotted line shows the profile of the optical lattice.}\label{prof}
\end{figure}

In the loading scenario discussed above, the key idea was to replace
suddenly the harmonic trap in the longitudinal direction with a
shallow periodic potential of a one-dimensional optical lattice; in
this case, the final state corresponding to a gap soliton is
composed by a few hundred of atoms. Moreover, the matter-wave gap
solitons created in this way are very narrow, and they occupy only a
few lattice sites. In addition, the loading efficiency is relatively low.
For that reason, below we suggest and discuss in detail the other
method for generating matter-wave gap solitons. In this second
scenario, we propose replacing suddenly the longitudinal harmonic
trap with a deep lattice, {\it and then gradually reduce the lattice
depth to a certain final value}. This kind of the condensate
dynamics does not leave atoms as much freedom to escape the central
region as they had before, ergo we can achieve higher values of the
atomic densities in the soliton. Moreover, by applying this method
we can end up with very shallow traps, much below the depth
accessible in the first scenario. In Fig.~\ref{adibatic}, we present
the result of our numerical simulations of the one-dimensional model
for an adiabatic reduction of the lattice depth from $\varepsilon=8
E_{\mathrm{recoil}}$ to $\varepsilon=E_{\mathrm{recoil}}$. During
this dynamics, the size of the condensate grows considerably due to
a strong reduction of the lattice depth. In simulations shown in
Fig.~\ref{adibatic}, we were able to decrease the lattice depth to
a level 2.5 times lower than  in the first method discussed above,
and we still obtained a stable state with about 500 atoms. The profile
of the corresponding matter-wave gap soliton is presented in
Fig.~\ref{prof}, and the noticeable difference between
Fig.~\ref{profil} and Fig.~\ref{prof} is the width of the created
gap soliton. Both gap solitons contain the same total number of
atoms but, since in the second method we use much weaker lattice,
the atoms spread over much larger region.

\begin{figure}[tbp]
\includegraphics[width=8.5cm]{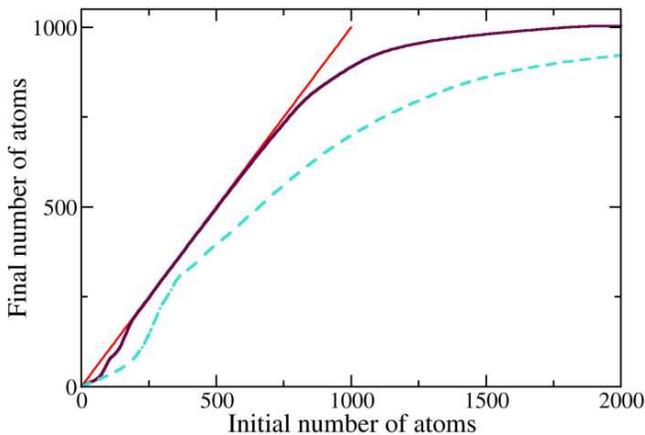}
\caption{(color online) Number of atoms in the window [$-100\mu$m,
$100\mu$m], which is approximately equal to the number of atoms in
the generated gap soliton, at $t=500$ms, as a function of the
initial number of atoms. Dashed line corresponds to the scenario
with a constant lattice depth of $\varepsilon=2.3
E_{\mathrm{recoil}}$, and the solid line-- to the case of the
adiabatic reduction of the lattice potential after switching it on
from $\varepsilon=8 E_{\mathrm{recoil}}$ to $\varepsilon=2.3
E_{\mathrm{recoil}}$ during the first $200$ms. Straight line marks
the available number of atoms.}\label{varN}
\end{figure}

To compare the efficiency of these two generation schemes,  we
perform additional numerical simulations and study how a variation
of the initial number of atoms and the lattice depth influence the
final number of atoms in the matter-wave gap soliton. In
Fig.~\ref{varN}, we monitor the efficiency of the soliton generation
by counting the atoms that remain after $t=0.5$s within the window
of 200 microns around the central peak. This choice is somehow
arbitrary, but the result does not depend qualitatively on the size
of the window. We observe that our adiabatic scheme allows to
achieve almost 100\% generation efficiency in a broad range of the
initial atom number. We also reveal that there exists a limit in the
final number of atoms that we can trap into the gap soliton. Both
the schemes display saturation above 1000 atoms. We can estimate the
maximal number of atoms that can be trapped in the soliton by
employing the effective-mass approximation and assuming that the
size of the soliton can not be smaller than one lattice well. Using
the parameters given above we estimate the upper limit to be around
1200 atoms, that is by almost five times larger that that reported
in the first experimental observation~\cite{Oberthaler}.

\begin{figure}[tbp]
\includegraphics[width=8.5cm]{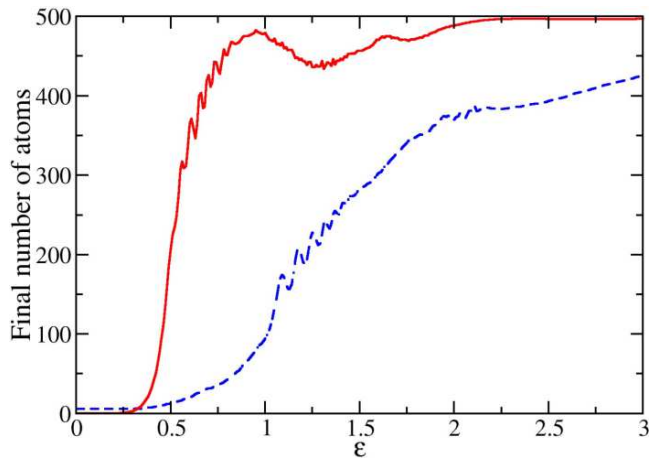}
\caption{(color online) Number of atoms in the window [$-100\mu$m,
$100\mu$m] at $t=5$s which is approximately equal to the number of
atoms in the generated gap soliton as a function of the final depth
of the optical lattice. Initially, there are 500 atoms in the
condensate. Dashed line corresponds to the case of a constant
lattice depth, and the solid line-- to the case of the adiabatic
reduction of the lattice potential from $\varepsilon=8
E_{\mathrm{recoil}}$ to the final depth during the first $500$\,ms.}
\label{vareps}
\end{figure}

It is clear that the adiabatic scheme enables to achieve higher
final population, and Fig.~\ref{vareps} shows clearly the advantage
of this method. In this figure, we plot the final number of atoms in
the gap soliton (we use the similar criterion as discussed above,
starting with 500 atoms) as a function of the final value of the
optical lattice depth. The adiabatic method leads to much smaller
loss of atoms during the loading process, and it can be used to
generate matter-wave gap solitons in much shallower lattices.
Solitons generated in this way span over quite large number of
lattice sites, and they exhibit much higher mobility, which can be
useful for potential applications, see ~\cite{Beata}. 
The latter feature that can be
seen in Fig.~\ref{vareps}, which shows a clear threshold for the
soliton generation better pronounced for the adiabatic approach.

Finally, we comment briefly on the soliton lifetime. In our model
and numerical simulations we neglect inelastic losses, so that the
timescale at which we follow the soliton generation dynamics remains
quite long. Still, we have checked that including inelastic losses
does not change significantly our main results, due to low densities
of the localized structures generated, but it remains the main
factor that will limit the lifetime of the gap solitons in real
experiments.

\section{Coupled-mode theory of self-trapping}

In this section, we introduce a simple analytical model which allows
to get a deeper insight into the physics of the soliton generation
and self-trapping in the nonlinear systems with repulsive
interaction. Such a model describes the dynamics in the momentum
space, but it is restricted by the first Brillouin zone of the
bandgap spectrum.

To derive the model, we decompose the condensate wavefunction in the
basis of the Bloch functions and then show the spontaneous migration
towards the region of the negative effective mass.

Before the derivation, we introduce the dimensionless variables
$z'=z/d$, $t'=t \omega$, $\Psi'=\Psi \sqrt{d/N}$,
where $\omega=\hbar/(m d^2)$ and $N$ is the number of atoms,
and define $g=4 \pi d N a_{\mathrm{1D}}$ and
$\varepsilon'=\varepsilon/(\hbar\omega)$.

First, we omit the primes for simplicity and apply the Fourier
transform of the one-dimensional rescaled model
\begin{eqnarray}
i \frac{\partial \Psi(k, t)}{\partial t} = \frac{k^2}{2} \Psi(k, t)
-\frac{\varepsilon}{4}[\Psi(k - 2\pi, t) +
\Psi(k + 2\pi, t)] + && \nonumber \\
+\frac{g}{2\pi}\int \int_{-\infty}^{+\infty}\Psi(k_1, t) \Psi^{*}(k_2, t)
\Psi(k-k_1+k_2, t) dk_1 dk_2. \nonumber
\end{eqnarray}
For the first Brillouin zone, i.e. $|k^{\prime}| < \pi$,  we define
the Bloch functions $u(k,k')$ of energy $E(k')$ satisfying
\begin{displaymath}
 E(k') u(k, k')  =
\frac{k^2}{2} u(k, k') \\ -\frac{\varepsilon}{4}[u(k - 2\pi, k') +
u(k + 2\pi, k')],
\end{displaymath}
with normalization condition $\int u(k,k'_1) u^{*}(k,k'_2) dk =
\delta(k'_1-k'_2)$. According to our assumption mentioned above we
can decompose the wavefunction $\Psi(k,t)$ into these states
\begin{displaymath}
 \Psi(k, t)  = \int_{-\pi}^{\pi} \Phi(k',t) u(k,k') dk',
\end{displaymath}
and obtain the equation for the amplitudes $ \Phi(k',t)$
\begin{eqnarray*}
&i&\!\!\!\! \frac{\partial \Phi(k', t)}{\partial t} = E(k') \Phi(k', t) + \\
&+&\frac{g}{2\pi} \int \int \int_{-\pi}^{\pi}
\Phi(k'_1, t) \Phi^{*}(k'_2, t) \Phi(k'_3, t) dk'_1 dk'_2 dk'_3 \times  \\
&\times& \int \int \int_{-\infty}^{+\infty}
u^{*}(k_0,k') u(k_1,k'_1) u^{*}(k_2,k'_2) \times \\
& \times& u(k_0 - k_1 + k_2,k'_3) dk_0 dk_1 dk_2 .
\end{eqnarray*}
For a weak lattice we assume that $u(k,k') \approx \delta(k-k')$;
this approximation works well everywhere except the regions close to
the band edge. Using this assumption, we obtain
\begin{eqnarray}
&&\!\!\!\!i \frac{\partial \Phi(k, t)}{\partial t} = E(k) \Phi(k, t) + \label{ebloch} \\
&&\!\!\!\!+\frac{g}{2\pi}\int\int_{-\pi}^{\pi} \Phi(k_1, t) \Phi^{*}(k_2, t) \Phi(k - k_1
+ k_2, t) dk_1 dk_2. \nonumber
\end{eqnarray}
We further simplify the dynamics by introducing a two-mode
approximation. Namely, we assume that to the first order our
wavefunction consists of the constant and modulated parts,
$\Phi(k,t)=[a(t) +  \sqrt{2} b(t) \cos(k)]/\sqrt{2\pi}$, $E(k) = E_0 +
E_1 \cos(k)$, where $E_1 < 0$ and $|a|^2+|b|^2=1$. Equation
(\ref{ebloch}) yields
\begin{eqnarray}
\frac{da}{dt} &=& -\frac{i}{\sqrt{2}} E_1 b - i g |a|^2 a, \nonumber\\
\frac{db}{dt} &=& -\frac{i}{\sqrt{2}} E_1 a - \frac{i}{2} g |b|^2 b. \label{dotab}
\end{eqnarray}
After substituting
\begin{eqnarray*}
a(t) &=& \sqrt{\frac{1-z(t)}{2}} \exp [i \phi_0(t)],\\
b(t) &=& \sqrt{\frac{1+z(t)}{2}} \exp [i (\phi_0(t) - \phi(t))],\\
\end{eqnarray*}
we obtain the asymmetric Josephson junction equations
\begin{eqnarray*}
\frac{d \phi}{dt} &=& \frac{z}{\sqrt{1 - z^2}} \cos \phi + \Lambda \left(z - \frac{1}{3}\right), \label{phidot}\\
\frac{d z}{d t} &=& -\sqrt{1 - z^2} \sin \phi, \label{zdot}
\end{eqnarray*}
where we introduce $\Lambda=3 g /(4\sqrt{2} |E_1|)$, and use
rescaled time $t_{\rm new} = \sqrt{2} |E_1| t$. To describe the
migration of waves from the region of normal to anomalous
diffraction, we introduce the quantity $M(t)$, which stands for a
difference between the population of atoms in the normal (central)
and anomalous (outer) regions of the Brillouin zone
\begin{eqnarray*}
M(t) &=& \int_{-\pi/2}^{\pi/2} |\Phi|^2 dk -
(\int_{-\pi}^{-\pi/2} |\Phi|^2 dk + \int_{\pi/2}^{\pi} |\Phi|^2 dk)= \nonumber \\
&=& \frac{4 \sqrt{2}}{\pi}\, \mathrm{Re} (a b^*)= \frac{2 \sqrt{2}}{\pi} \sqrt{1-z^2} \cos \phi.
\end{eqnarray*}
If $M>0$, the condensate is composed mostly of the matter waves from
the normal diffraction region, and for $M<0$ -- of the matter waves
from the anomalous diffraction region. The temporal derivative of
$M$ can be expressed as
\begin{eqnarray}
\frac{dM}{dt} &=& \frac{2
\sqrt{2}}{\pi}\Lambda\left(z-\frac{1}{3}\right)
\frac{dz}{dt}.\label{dotM}
\end{eqnarray}

\begin{figure}[tbp]
\includegraphics[width=8.5cm]{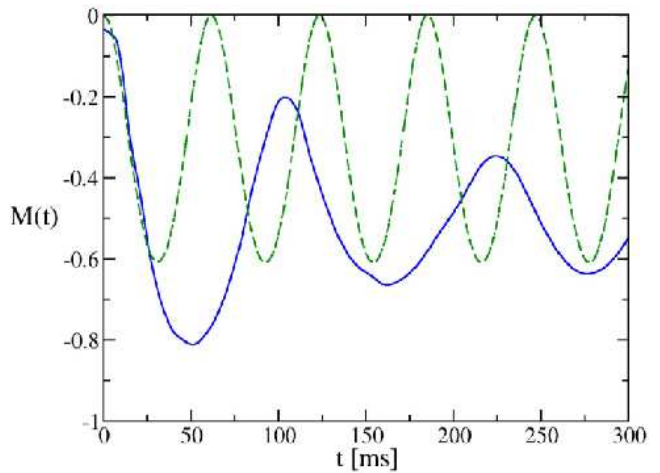}
\caption{(color online) Evolution of the parameter $M(t)$ above the
threshold for soliton formation ($N=500$ atoms), corresponding to
Fig.~\ref{columns}(c). Solid line shows the results of the
one-dimensional GPE model, and dashed line shows results of the
coupled-mode model, Eqs.~(\ref{dotab}).}\label{m500}
\end{figure}

\begin{figure}[tbp]
\includegraphics[width=8.5cm]{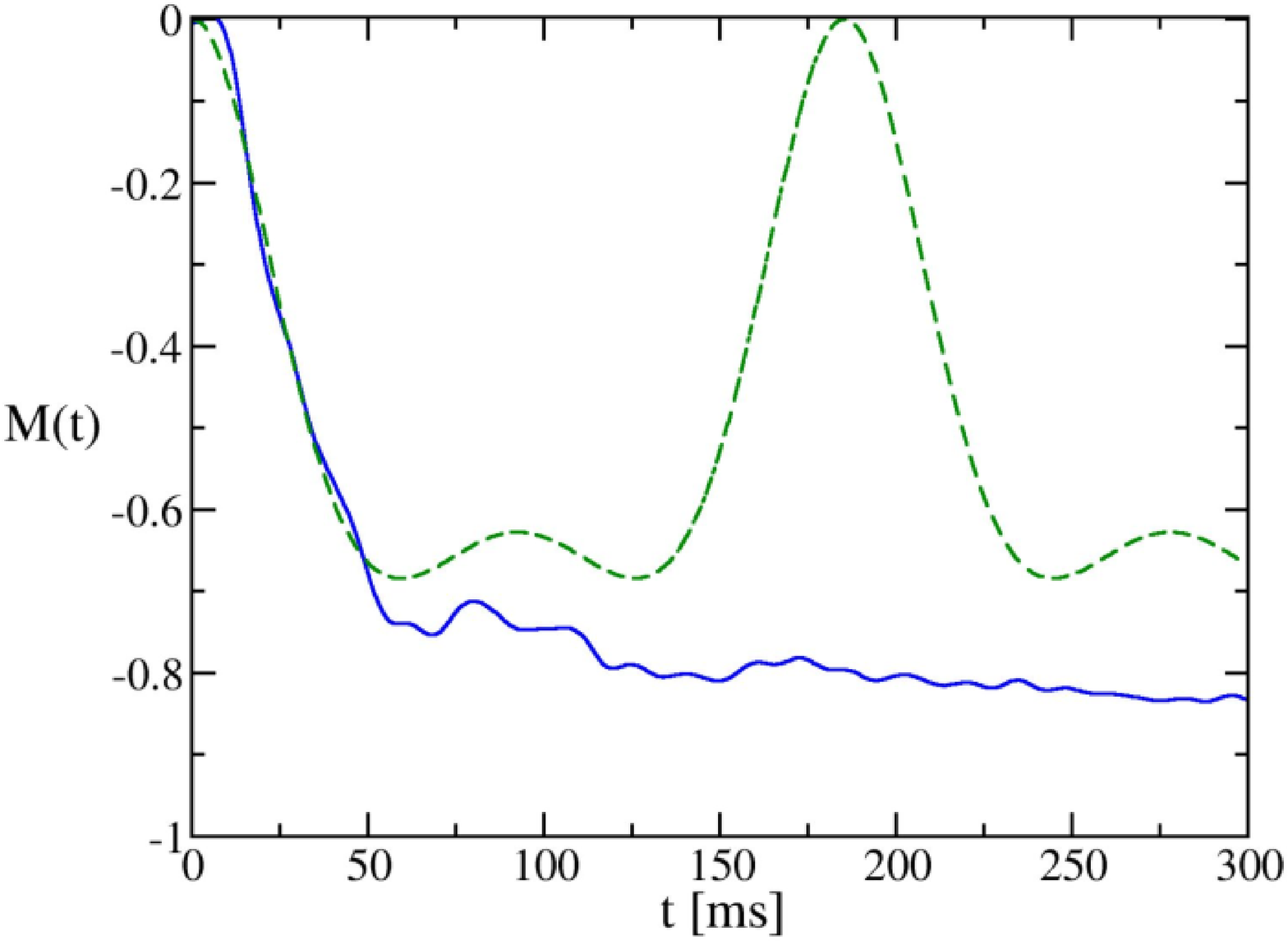}
\caption{(color online) Same as in Fig.~\ref{m500}, but in the case
of $N=200$ atoms, see Fig.~\ref{columns}(b).}\label{m200}
\end{figure}

To apply this equation for describing the dynamics of the soliton
generation, we assume that at $t=0$ the condensate is in the ground
state of a single lattice site. In this state, the first band is
populated approximately uniformly with a constant phase, so that
$a=1$ and $b=0$ and, consequently, $z=-1$ and $M(0)=0$. Since $z$
cannot be smaller than $-1$, $z$ will grow in time and from
Eq.~(\ref{dotM}) we see that $M(t)$ will initially decrease for a
repulsive condensate ($\Lambda>0$). Hence the matter waves will
change their state migrating to the regions of the negative
effective mass, allowing the gap soliton to be formed.

This general picture has been confirmed by direct numerical
simulations of both the full Gross-Pitaevskii equation and our
simplified coupled-mode model, as shown in Fig.~\ref{m500} and
Fig.~\ref{m200}. In Fig.~\ref{m500} we present a comparison of the
parameter $M(t)$ calculated from one-dimensional GPE and
Eq.~(\ref{dotab}) for the case when the soliton is formed (500 atoms
in the initial condensate) and Fig.~\ref{m200} corresponds to the
case below the threshold of the gap-soliton formation. In both cases
the initial dynamics is similar and well described by our simple
model. However, the nonlinear dynamics observed in the simulations
of the full model is more complicated because the energy exchange
between the two major modes is accompanied by excitation of other
modes in the system not taken into account by our simplified
two-mode approximation. In the results obtained for the full model,
this effect is observed as an effective damping of the oscillations.

\section{Conclusions}

We have revealed that the generation of matter-wave gap solitons can
be easier than expected. We propose and demonstrate numerically two
generation schemes in which a robust, long-lived stationary
wavepacket in the form of a matter-wave gap soliton is created in a
repulsive Bose-Einstein condensate placed into a one-dimensional
optical lattice. The suggested generation method looks simple and
efficient, and it relies on a relaxation of the initial distribution
of atoms to the appropriate soliton state. For this scheme, lifetime
of the final state is limited only by the lifetime of the
condensate. We presented a simple theoretical model to illustrate
how the generation method works.

\section*{Acknowledgements}

M.M. acknowledges a support from the KBN grant 2P03 B4325 and
the Foundation for Polish Science. M.T. was
supported by the Polish Ministry of Scientific Research and
Information Technology under grant PBZ MIN-008/P03/2003. The work of
Y.K. has been supported by the Australian Research Council through
the Centre of Excellence Program; Y.K. thanks Elena Ostrovskaya and
Tristram Alexander for useful discussions.

\end{document}